\documentclass[a4paper,twocolumn,prb,superscriptaddress,showpacs,tight]{revtex4-1}
\usepackage{ae,aecompl}
\usepackage[T1]{fontenc}
\usepackage[T1]{fontenc}
\usepackage[latin1]{inputenc}
\usepackage{bm}
\usepackage{amsmath}
\usepackage{amssymb}
\usepackage{graphicx}
\usepackage{esint}

\makeatletter

\pdfoutput=1
\@ifundefined{definecolor}
 {\usepackage{color}}{}
\usepackage[pdftex]{hyperref}
\usepackage{dcolumn}
\usepackage{bm}
\usepackage{times}

\makeatother

\begin{document}

\title{Efficient graphene-based photodetector with two cavities}

\author{Aires Ferreira}

\affiliation{ Graphene Research Centre and Department of Physics, National
University of Singapore, 2 Science Drive 3, Singapore 117542}

\affiliation{ Department of Physics and Center of Physics, University of
Minho, P-4710-057, Braga, Portugal}

\author{N. M. R. Peres}

\affiliation{ Graphene Research Centre and Department of Physics, National
University of Singapore, 2 Science Drive 3, Singapore 117542}

\affiliation{ Department of Physics and Center of Physics, University of
Minho, P-4710-057, Braga, Portugal}

\author{R. M. Ribeiro}

\affiliation{ Department of Physics and Center of Physics, University of
Minho, P-4710-057, Braga, Portugal}

\author{T. Stauber}

\affiliation{ Department of Physics and Center of Physics, University of
Minho, P-4710-057, Braga, Portugal}

\affiliation{Departamento de Física de la Materia Condensada and Instituto
Nicolás Cabrera, Universidad Autónoma de Madrid, E-28049 Madrid, Spain }

\date{\today}
\begin{abstract}
We present an efficient graphene-based photodetector with two Fabri-P\'erot
cavities. It is shown that the absorption can reach almost 100\% around
a given frequency, which is determined by the two-cavity lengths. It is also shown
that  hysteresis in the absorbance is possible, with the transmittance
amplitude of the mirrors working as an external driving field. The
role of non-linear contributions to the optical susceptibility of
graphene is discussed. 
\end{abstract}


\maketitle

\section{Introduction}

\label{sec_introd}

Exploring the optical properties of graphene\cite{rmp,nair,rmpperes}
for photodetection is one of the most promising applications of graphene.\cite{Avouris}
Graphene has no gap and its conductivity is essentially independent
of frequency\cite{nmrPRB06,abergel,gusyninIJMPB,falkov1,falkov2}
for photon energies up to 2~eV. These properties, combined with the
intrinsic chemical and mechanical stability of graphene, pave the way
for broad band optoelectronics.

Depending on the authors, graphene is characterized as presenting
remarkable high absorption\cite{Abajo} or weak absorption.\cite{Novoselov}
Indisputable however is the fact that pristine graphene absorbs
about 2.3\% of the light impinging on it. A considerable frequency
dependence of the absorption appears as the photon energy approaches
$\sim4.7$~eV, due to the combined effects of the Van Hove singularity
in graphene's electronic $\pi-$spectrum and excitonic many-body effects.\cite{LiLouie}
The value of 2.3\% can also be interpreted as the probability for
photon absorption in a single passing through the material.

To enhance the absorption of graphene several mechanisms have been
proposed, ranging from hybrid materials, containing carbon, boron,
and nitrogen,\cite{Novoselov,NewDirections} nano-patterning
of a graphene sheet\cite{FullAbajo},
to strain engineering,\cite{Novoselov,NewDirections} and plasmonics.\cite{Ozbay,EPLPeres}
In the latter case, micro-sized ribbons patterned in a single graphene
sheet,\cite{Zettl} and metallic arrays on top of graphene,\cite{Geim}
lead to an enhancement of the near field, thus increasing light absorption
and producing larger photo-currents compared to the case of pristine graphene. 
Current plasmonics-based approaches
are limited to specific spectral bands. 
In what concerns hybrid
materials for photonic applications, there is still a long way to go before these
become possible.\cite{Novoselov,NewDirections}

Photodetection, depending on the type of application, may require
efficient absorption of light in a narrow spectral band. Then, it
is conceivable to explore Fabry-P\'erot interference for producing an
efficient photodetector tailored for a specific application. The concept
is  simple: in a Fabry-P\'erot interferometer a photon may be
trapped inside the cavity for a \textit{long time}, undergoing many
round trips before leaving it; indeed, if a material able to absorb photons 
is introduced inside the optical cavity, without
significantly changing the cavity's finesse, as is the case of graphene,
most of the 
photons of the right frequency entering  the optical cavity will be absorbed. 

For a combined cavity-graphene system, it is
important to quantity the magnitude of non-linear optical effects,
ensuring whether linear response theory can be used for describing
the absorption process inside the cavity. Thus, in the present work,
we discuss the non-linear optical susceptibility of graphene first, 
presenting later the single and double cavity photodetectors.

This article is organized as follows: in Sec.~\ref{sec_bloch} we
introduce Bloch's equations for graphene and compute both the linear
and non-linear optical susceptibility. As aforementioned, the calculation
of the non-linear part is essential for a critical analysis of the
optical response of graphene. In Sec.~\ref{sec_power} we give the
power series solution of Bloch's equation and discuss the validity
of perturbation theory. In Sec.~\ref{sec_cavity} we introduce
the mathematical description of a graphene-based photodetector. Having
shown that non-linear contributions to the optical susceptibility
are relevant only at very high field intensities, we employ linear
response theory to describe the graphene-based photodetector with
two coupled optical cavities.

\section{Derivation of Bloch equations of motion\label{sec_bloch}}

The calculation of the optical properties of a given material can
be obtained from the solution of Bloch's equations.\cite{Haug,Malic,Peres_Excitonic,EOM}
Below, we derive  Bloch's  differential equations for graphene and
give its solution for an incoming electromagnetic plane wave impinging
on the material.

The Hamiltonian of the electrons in graphene, in the presence of an
electromagnetic field, reads (spin index implicit):

\begin{eqnarray}
H & = & \sum_{\bm{k}}E_{c}(\bm{k})a_{c,\bm{k}}^{\dag}a_{c,\bm{k}}+E_{v}(\bm{k})a_{v,\bm{k}}^{\dag}a_{v,\bm{k}}\nonumber \\
 & + & v_{F}eA(t)\sum_{\bm{k}}(d_{cv,\bm{k}}^{x}a_{c,\bm{k}}^{\dag}a_{v,\bm{k}}+d_{vc,\bm{k}}^{x}a_{v,\bm{k}}^{\dag}a_{c,\bm{k}})\,,\label{eq:hamilt}
\end{eqnarray}
where $E_{c/v}(\bm{k})=\pm v_{F}\hbar k$, $v_{F}=3ta_{0}/\hbar$
is the Fermi velocity ($t\simeq2.7$ eV and $a_{0}=1.4$ \AA{}\,
are the hoping integral and the carbon-carbon distance, respectively),
$a_{c/v,\bm{k}}^{\dag}$ is the creation operator of an electron with
wave number $\bm{k}$ in the conduction/valence band, 
\begin{equation}
d_{vc,k}^{x}=-i\sin\theta
\end{equation}
is the matrix element of the dipole operator ($d_{cv,k}^{x}=i\sin\theta$),
$A(t)=A_{0}\sin(\omega t)$ is the vector potential for a linearly
polarized electromagnetic plane wave, $e>0$ is the elementary charge,
$\omega$ is the frequency of light, and 
\begin{equation}
\theta=\arctan\frac{k_{y}}{k_{x}}\,.
\end{equation}
Since we have $E(t)=-\frac{\partial A(t)}{\partial t}=-\omega A_{0}\cos(\omega t)\,,$
we can write $A(t)$ as 
\begin{equation}
A(t)=i\frac{E_{0}}{2\omega}e^{-i\omega t}+\frac{E_{0}}{2i\omega}e^{i\omega t}\,,
\end{equation}
 where $E_{0}$ is the intensity of the electric field. 

At the heart of the present approach is  Heisenberg's equation of motion for
the polarization operator $\hat{P}_{vc,\bm{k}}\equiv a_{v,\bm{k}}^{\dag}a_{c,\bm{k}}$,
namely, 
\begin{equation}
-i\hbar\frac{d\hat{P}_{vc,\bm{k}}}{dt}=[H,\hat{P}_{vc,\bm{k}}]\,.
\end{equation}
 The explicit form of the equation of motion is 
\begin{eqnarray}
-i\hbar\frac{d\hat{P}_{vc,\bm{k}}}{dt} & - & i\hbar\gamma_{2,\bm{k}}\hat{P}_{vc,\bm{k}}=-[E_{c}(\bm{k})-E_{v}(\bm{k})]\hat{P}_{vc,\bm{k}}\nonumber \\
 & + & v_{F}eA(t)(a_{c,\bm{k}}^{\dag}a_{c,\bm{k}}-a_{v,\bm{k}}^{\dag}a_{v,\bm{k}})d_{cv,\bm{k}}^{x}\,.
\end{eqnarray}
In the above, $\gamma_{2,\bm{k}}$ is the phenomenological relaxation
rate of the polarization. In addition, we need the equation of motion
for the number operator, $\hat{n}_{\lambda,\bm{k}}$, both in the
conduction and valence bands, reading 
\begin{eqnarray}
-i\hbar\frac{d\hat{n}_{c,\bm{k}}}{dt}-i\hbar\gamma_{1,\bm{k}}\hat{n}_{c,\bm{k}} & = & ev_{F}A(t)D_{vc,\bm{k}}\,,\label{eq_nc}\\
-i\hbar\frac{\partial\hat{n}_{v,\bm{k}}}{\partial t}-i\hbar\gamma_{1,\bm{k}}\hat{n}_{v,\bm{k}} & = & ev_{F}A(t)D_{vc,\bm{k}}^{\ast}\,,\label{eq_nv}
\end{eqnarray}
 where 
\begin{equation}
D_{vc,\bm{k}}=d_{vc}^{x}\hat{P}_{vc,\bm{k}}-d_{cv}^{x}\hat{P}_{cv,\bm{k}}\,,
\end{equation}
and $\gamma_{1,\bm{k}}$ is the phenomenological relaxation rate for
the occupation number of a state $\bm{k}$. It is convenient to define
the population operator, $\hat{N}_{\bm{k}}\equiv\hat{n}_{c,\bm{k}}-\hat{n}_{v,\bm{k}}$,
whose average obeys the following differential equation 
\begin{equation}
\hbar\frac{d}{dt}\hat{N}_{\bm{k}}+\hbar\gamma_{1,\bm{k}}
\hat{N}_{\bm{k}}=2ev_{F}\sin\theta A(t)(\hat{P}_{vc,\bm{k}}+\hat{P}_{cv,\bm{k}})\,.
\end{equation}
We denote the average of $\hat{N}_{\bm{k}}$, $\hat{P}_{vc,\bm{k}}$,
and $\hat{P}_{cv,\bm{k}}$, by $N_{\bm{k}}$, $P_{vc,\bm{k}}$, and
$P_{cv,\bm{k}}$, respectively. The latter averages obey the following
set of linear, first order, differential equations (known as Bloch's
equations): 
\begin{eqnarray}
\frac{d}{dt}P_{vc,\bm{k}}+\gamma_{2,\bm{k}}P_{vc,\bm{k}} & = & -i\epsilon_{\bm{k}}P_{vc,\bm{k}}-d_{\bm{k}}A(t)N_{\bm{k}}\,,\label{eq_P_av_dif}\\
\frac{d}{dt}P_{cv,\bm{k}}+\gamma_{2,\bm{k}}P_{cv,\bm{k}} & = & i\epsilon_{\bm{k}}P_{cv,\bm{k}}-d_{\bm{k}}A(t)N_{\bm{k}}\,,\\
\frac{d}{dt}N_{\bm{k}}+\gamma_{1,\bm{k}}N_{\bm{k}} & = & 2d_{\bm{k}}A(t)(P_{vc,\bm{k}}+P_{cv,\bm{k}})\,,\label{eq_N_av_dif}
\end{eqnarray}
 where $d_{\bm{k}}=v_{F}e\sin\theta/\hbar$ and $\epsilon_{\bm{k}}=2v_{F}k$.
We note that the differential equation for $P_{cv,\bm{k}}$ is redundant,
since $\hat{P}_{cv,\bm{k}}=[\hat{P}_{vc,\bm{k}}]^{\dag}$. The differential
equations are solved together with the initial conditions ($t=-\infty$.):
$P_{vc,\bm{k}}=0$ and $N_{\bm{k}}=n_{0}=f[E_{c}(\bm{k})]-f[E_{v}(\bm{k})]$,
where $f(x)$ is the equilibrium Fermi distribution. The condition
$n_{0}=-1$ applies to neutral graphene at zero temperature.

We note in passing that in the absence of relaxation 
mechanisms, i.e.,~$\gamma_{1,\bm{k}}=\gamma_{2,\bm{k}}=0$,
the quantity $N_{\bm{k}}^{2}+4P_{vc,\bm{k}}P_{cv,\bm{k}}$ is a constant
of motion. We also note that there is no fundamental reason why $\gamma_{1,\bm{k}}$
should be equal to $\gamma_{2,\bm{k}}$, albeit they generally are
of the same order of magnitude; in order to simplify the mathematical
expressions, we assume below that $\gamma_{1,\bm{k}}=\gamma_{2,\bm{k}}\equiv\gamma$.
This procedure is justified since making $\gamma_{1,\bm{k}}\ne\gamma_{2,\bm{k}}$
does not alter the final qualitative conclusions.

\section{Power series solution to Bloch's equations\label{sec_power}}

In the present section we obtain the solution of Bloch's equations.
It is convenient to employ the shorthand notation, $P_{vc,\bm{k}}=X(t)\equiv X$,
$N_{\bm{k}}=N(t)\equiv N$, $\epsilon_{\bm{k}}=\epsilon$, and $d_{\bm{k}}A(t)=a(t)$.
Using this notation, Bloch's equations have the form 
\begin{eqnarray}
\frac{d}{dt}X+\gamma X & = & -i\epsilon X-a(t)N\,,\label{eq_Bloch_X}\\
\frac{d}{dt}N+\gamma N & = & 2a(t)(X+X^{\ast})=4a(t)\Re X\,.\label{eq_Bloch_N}
\end{eqnarray}
We now assume a power series solution for $X$ and $N$\cite{Schafer2002}
\begin{eqnarray}
 &  & a(t)\rightarrow\lambda a(t)\,,\label{eq_a_lamd}\\
 &  & X(t)\rightarrow\lambda x_{1}(t)+\lambda^{2}x_{2}(t)+\lambda^{3}x_{3}(t)+\ldots\,,\label{eq_x_lamd}\\
 &  & N(t)\rightarrow n_{0}+\lambda n_{1}(t)+\lambda^{2}n_{2}(t)+\lambda^{3}n_{3}(t)+\ldots\,,\label{eq_n_lamd}
\end{eqnarray}
 where $\lambda$ is a bookkeeping of the power of the amplitude of
the electric field; it is useful to use the notation $n_{\bm{k}}\equiv n_{0}$.
Introducing the series expansions (\ref{eq_a_lamd}), (\ref{eq_x_lamd}),
and (\ref{eq_n_lamd}) in Eqs.~(\ref{eq_Bloch_X}) and (\ref{eq_Bloch_N})
it is simple to see that only  odd powers, $x_{2m+1}$, of the
polarization are non-zero, whereas for the population only  even
powers, $n_{2m}$, are finite, with $m$ an integer number, including
zero.

Two relevant dimensionless parameters involving the intensity of the
incoming field, ${\cal W}_{i}$, are: 
\begin{equation}
\beta_{\gamma}=\pi\alpha\frac{27}{4}\frac{{\cal W}_{i}a_{0}^{2}t^{2}}{\hbar^{3}\omega^{2}\gamma^{2}}\,,
\end{equation}
 and 
\begin{equation}
\beta_{\omega}=\pi\alpha\frac{27}{4}\frac{{\cal W}_{i}a_{0}^{2}t^{2}}{\hbar^{3}\omega^{2}(\omega^{2}+\gamma^{2}/2)}\approx\pi\alpha\frac{27}{4}\frac{{\cal W}_{i}a_{0}^{2}t^{2}}{\hbar^{3}\omega^{4}}\,,
\end{equation}
 where ${\cal W}_{i}=E_{0}^{2}\epsilon_{0}c/2$, $\alpha=e^{2}/(4\pi\epsilon_{0}\hbar c)$
is the fine structure constant, and we also have assumed $\omega\gg\gamma$
in $\beta_{\omega}$. When either $\beta_{\gamma}>1$ or $\beta_{\omega}>1$,
the perturbative solution breaks down and the full series has to be
resumed. The choice of prefactors in $\beta_{\gamma}$ and $\beta_{\omega}$
will be apparent later in the text. Numerically, the intensities setting
the limit of validity of perturbation theory are 
\begin{equation}
{\cal W}_{i,\gamma}=(\hbar\omega)^{2}(\hbar\gamma)^{2}\times10^{3}\frac{{\rm {GW}}}{{\rm {cm^{2}}}}
\end{equation}
 from $\beta_{\gamma}$=1, and 
\begin{equation}
{\cal W}_{i,\omega}=(\hbar\omega)^{4}\times10^{3}\frac{{\rm {GW}}}{{\rm {cm^{2}}}}
\end{equation}
 from $\beta_{\omega}$=1, with $\hbar\omega$ and $\hbar\gamma$
expressed in electron-volt; for graphene we have $\hbar\gamma\sim10$~meV.
Taking a representative value of $\hbar\omega\sim0.5$ eV, we obtain
\begin{eqnarray}
{\cal W}_{i,\gamma} & \simeq & 2.5\times10^{-2}\frac{{\rm {GW}}}{{\rm {cm^{2}}}}\,,\\
{\cal W}_{i,\omega} & \simeq & 60\frac{{\rm {GW}}}{{\rm {cm^{2}}}}\,.
\end{eqnarray}
 It should be noted the three orders of magnitude difference between
the two cases.

\subsection{Linear optical susceptibility\label{sec_lin_sus}}

The calculation of the optical susceptibility of graphene can be made
for any frequency value. \cite{stauberFullSigma,LiLouie,RicardoStrain}
On the other hand, Hamiltonian (\ref{eq:hamilt}) is valid up to energies
of the order of $1$~eV, which translates into photon frequencies
of the order of $2$~eV. Hence, both for illustrating the method
and describing how the photodetector works, the Dirac cone approximation
suffices for our purposes.

The solution for the linear polarization, $x_{1}(t)$, is obtained
from 
\begin{equation}
\dot{x}_{1}+(i\epsilon+\gamma)x_{1}=-a(t)n_{0}\,,
\end{equation}
 which is easily solved by the integrating factor $e^{(i\epsilon+\gamma)t}$,
leading to
\begin{equation}
x_{1}(t)=-e^{-(i\epsilon+\gamma)t}\int_{-\infty}^{t}e^{(i\epsilon+\gamma)t'}a(t')n_{0}dt'\,.
\end{equation}
 In the particular case where $A(t)$ is described by a sinusoidal
function we obtain 
\begin{equation}
x_{1}(t)=-n_{\bm{k}}d_{\bm{k}}\frac{E_{0}}{2\omega}\left(\frac{e^{i\omega t}}{\omega+\epsilon-i\gamma}+\frac{e^{-i\omega t}}{\omega-\epsilon+i\gamma}\right)\,.
\end{equation}
 In general, the total polarization is computed from 
\begin{eqnarray}
P_{x} & = & -ev_{F}g_{s}g_{v}\sum_{\bm{k}}(P_{vc,\bm{k}}d_{vc}^{x}+P_{cv,\bm{k}}d_{cv}^{x})\nonumber \\
 & = & \lambda P_{x,1}+\lambda^{3}P_{x,3}+\ldots\,,
\end{eqnarray}
 where $g_{s}$ and $g_{v}$ are the spin and valley degeneracy, respectively.
Recalling that $P_{vc,\bm{k}}=X$ and $P_{cv,\bm{k}}=X^{\ast}$, we
have, to first order in the electric field amplitude ($E_{0}=A_{0}/i\omega$)
\begin{eqnarray}
P_{x,1} & = & iev_{F}g_{s}g_{v}\sum_{\bm{k}}\sin\theta[x_{1}(t)-x_{1}^{\ast}(t)]\nonumber \\
 & = & \frac{E_{0}}{2}\chi_{1}(\omega)e^{-i\omega t}+\frac{E_{0}}{2}\chi_{1}(-\omega)e^{i\omega t}\,,\label{eq_Px1}
\end{eqnarray}
 with $\chi_{1}(-\omega)=\chi_{1}^{\ast}(\omega)$. Considering the
case of neutral graphene at zero temperature, we obtain for the real
part of the optical susceptibility, $\Re\chi_{1}(\omega)\equiv\chi_{1}'$,
the well known value 
\begin{equation}
\chi_{1}'=\frac{\pi e^{2}}{2h}\equiv\sigma_{0}\,,\label{eq_universal}
\end{equation}
dubbed the universal conductivity of graphene.\cite{nair} Given Eq.~(\ref{eq_universal}),
the imaginary part of the optical susceptibility reads $\Im\chi_{1}(\omega)\equiv\chi_{1}''=0$,
as follows from the Kramers-Kronig relation: 
\begin{equation}
\chi_{1}''(\omega)=-\frac{1}{\pi}{\cal P}\int_{-\infty}^{\infty}\frac{\chi_{1}'(x)}{x-\omega}dx\,.\label{eq_KK}
\end{equation}
Equivalent relations hold for non-linear response 
functions as well.\cite{Osamu_Nakada_1960,Boyd}

\subsection{Non-linear optical susceptibility\label{sec_nonlin_sus}}

To go beyond linear response, we have to compute how the population
changes relatively to its initial value when the field is turned on.
This amounts to compute $n_{2}(t)$. The latter can be obtained from
the solution of $x_{1}(t)$ according to 
\begin{equation}
n_{2}(t)=2e^{-\gamma t}\int_{-\infty}^{t}dt'a(t')[x_{1}(t')+x_{1}^{\ast}(t')]\,.
\end{equation}
 The occupancy second-order correction, $n_{2}(t)$, is a sum of two
contributions: $n_{2}(t)=n_{2a}+n_{2b}(t)$, where the first one is
independent of time. Explicitly we have 
\begin{equation}
n_{2a}=-n_{\bm{k}}d_{\bm{k}}^{2}\frac{E_{0}^{2}}{\omega^{2}}\left(\frac{1}{(\omega-\epsilon_{\bm{k}})^{2}+\gamma^{2}}+\frac{1}{(\omega+\epsilon_{\bm{k}})^{2}+\gamma^{2}}\right)\,,
\end{equation}
 and 
\begin{eqnarray}
n_{2b}(t) & = & -\frac{n_{\bm{k}}}{2}d_{\bm{k}}^{2}\frac{E_{0}^{2}}{\omega^{2}}\left[e^{-2i\omega t}\frac{2\omega-i\gamma}{4\omega^{2}+\gamma^{2}}\left(\frac{\omega-\epsilon_{\bm{k}}-i\gamma}{(\omega-\epsilon_{\bm{k}})^{2}+\gamma^{2}}\right.\right.\nonumber \\
 & + & \left.\left.\frac{\omega+\epsilon_{\bm{k}}-i\gamma}{(\omega+\epsilon_{\bm{k}})^{2}+\gamma^{2}}\right)+\mbox{c. c.}\right]\,.
\end{eqnarray}
 We note that $n_{2a}$ is a positive number, thus reducing the value
of $N(t)$ when the system is driven away from equilibrium.

Similarly, the calculation of $x_{3}(t)$ follows from 
\begin{equation}
x_{3}(t)=-e^{-(i\epsilon+\gamma)t}\int_{-\infty}^{t}e^{(i\epsilon+\gamma)t'}a(t')[n_{2a}+n_{2b}(t')]dt'\,.
\end{equation}
 The quantity $x_{3}(t)$ is a sum two different terms $x_{3}(t)=x_{3a}(t)+x_{3b}(t)$:
\begin{equation}
x_{3a}(t)=in_{\bm{k}}d_{\bm{k}}^{3}\frac{E_{0}^{3}}{4\omega^{3}}A(\epsilon_{\bm{k}},\omega)\frac{e^{i\omega t}}{i\omega+i\epsilon_{\bm{k}}+\gamma}+(\omega\rightarrow-\omega)\,,
\end{equation}
 where $(\omega\rightarrow-\omega)$ is obtained from the given explicit
term upon the replacement $\omega\rightarrow-\omega$, and 
\begin{equation}
x_{3b}(t)\approx-in_{\bm{k}}d_{\bm{k}}^{3}\frac{E_{0}^{3}}{4\omega^{3}}B(\epsilon_{\bm{k}},\omega)\frac{e^{i\omega t}}{i\omega+i\epsilon_{\bm{k}}+\gamma}+(\omega\rightarrow-\omega)\,.\label{eq_x3b}
\end{equation}
 Terms proportional to $e^{\pm i3\omega t}$ correspond to three photon
absorption and were neglected in Eq.~(\ref{eq_x3b}).\cite{Boyd}
We have also defined 
\begin{eqnarray}
A(\epsilon_{\bm{k}},\omega) & = & \frac{2}{(\omega-\epsilon_{\bm{k}})^{2}+\gamma^{2}}+\frac{2}{(\omega+\epsilon_{\bm{k}})^{2}+\gamma^{2}}\,,\\
B(\epsilon_{\bm{k}},\omega) & = & \frac{2\omega+i\gamma}{4\omega^{2}+\gamma^{2}}\left(\frac{\omega-\epsilon_{\bm{k}}+i\gamma}{(\omega-\epsilon_{\bm{k}})^{2}+\gamma^{2}}+\frac{\omega+\epsilon_{\bm{k}}+i\gamma}{(\omega+\epsilon_{\bm{k}})^{2}+\gamma^{2}}\right)\,.\nonumber \\
\end{eqnarray}
 It is important to note the symmetries, $A(\epsilon,\omega)=A(-\epsilon,\omega)$
and $B(\epsilon,\omega)=B(-\epsilon,\omega)$, which help in the calculation
of the total optical susceptibility.

\begin{figure}[ht]
\begin{centering}
\includegraphics[clip,width=8cm]{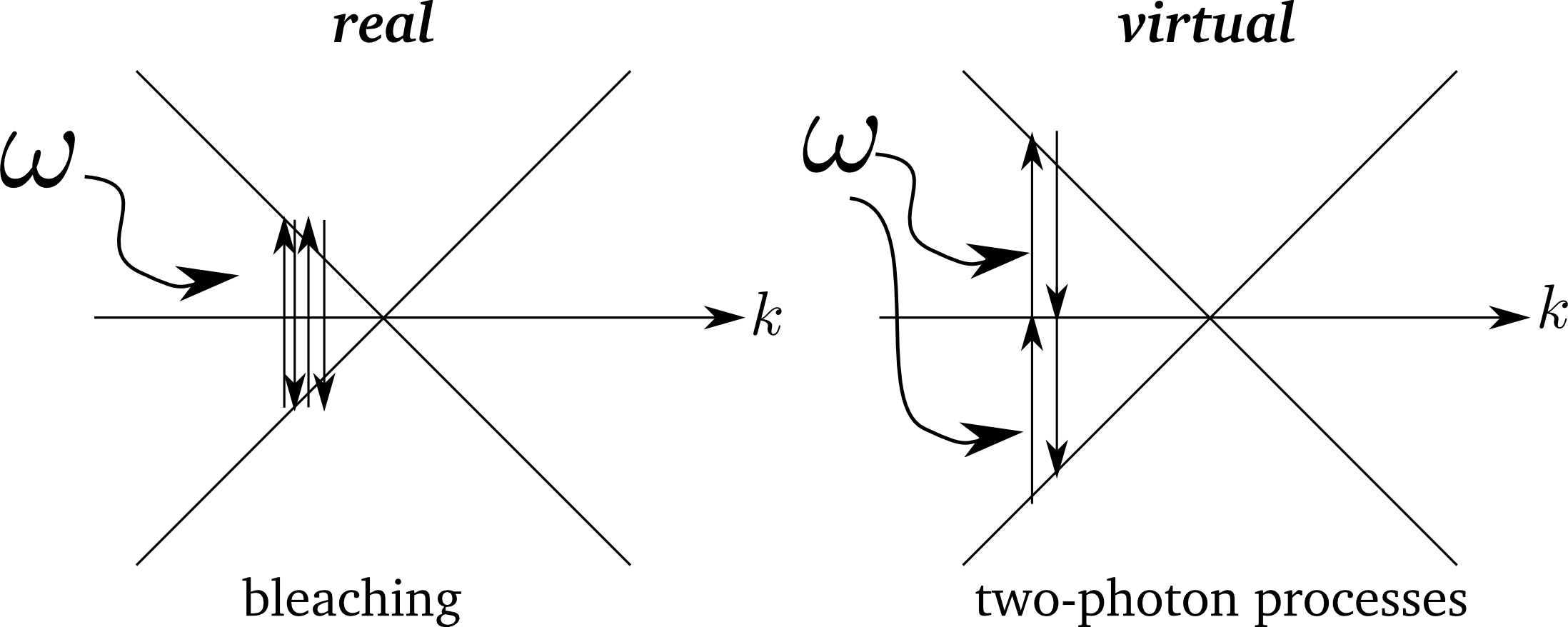} 
\par\end{centering}

\caption{\label{fig_two_photon}Third order optical processes in neutral graphene.
\textit{\emph{Bleaching}}: two photons are absorbed with no virtual
states involved. This process requires emission of a photon before
the second one is absorbed. Resonant two-photon processes: a process
where two photons are simultaneously absorbed involving a virtual
state. In the case represented here, that state is located at zero
energy. These type of processes also occur in traditional semiconductors,
when the photon energy is smaller than the band-gap.\cite{Boyd}}
\end{figure}

Analogously to $P_{x,1}$, the non-linear polarization $P_{x,3}$
is obtained from 
\begin{equation}
P_{x,3}=iev_{F}g_{s}g_{v}\sum_{\bm{k}}\sin\theta[x_{3}(t)-x_{3}^{\ast}(t)]\,.\label{eq_Px3}
\end{equation}
 Replacing the expression for $x_{3}(t)$ in Eq.~(\ref{eq_Px3})
we obtain (for neutral graphene at zero temperature) 
\begin{equation}
P_{x,3}=\chi_{3}^{(\omega;bl)}\frac{E_{0}}{2}e^{-i\omega t}+\chi_{3}^{(\omega;2\gamma)}\frac{E_{0}}{2}e^{-i\omega t}+\mbox{h. c.}\,,
\end{equation}
 where $\chi_{3}^{(\omega;bl)}$ and $\chi_{3}^{(\omega;2\gamma)}$
are given by 
\begin{eqnarray}
\chi_{3}^{(\omega;bl)} & = & -\sigma_{0}{\cal W}_{i}\pi\alpha\frac{3v_{F}^{2}}{\hbar\omega^{2}\gamma^{2}}=-\sigma_{0}\beta_{\gamma}\,,\\
\chi_{3}^{(\omega;2\gamma)} & = & -\sigma_{0}{\cal W}_{i}\pi\alpha\frac{3v_{F}^{2}}{\hbar\omega^{3}}\frac{2\omega}{4\omega^{2}+\gamma^{2}}=-\sigma_{0}\beta_{\omega}/2\,,
\end{eqnarray}
with the respective imaginary parts being negligible in the regime
$\hbar\gamma\ll\hbar\omega$. Note that both $\Re\chi_{3}^{(\omega;bl)}$
and $\Re\chi_{3}^{(\omega;2\gamma)}$ are negative due to saturate absorption. 
Also, both processes contribute to the imaginary part of the
refraction index of graphene. We should note that the limit $\gamma\rightarrow0$
can be taken in $\beta_{\omega}$ but not in $\beta_{\gamma}$. Indeed,
$\Re\chi_{3}^{(\omega;bl)}$ and $\Re\chi_{3}^{(\omega;2\gamma)}$
correspond to two different physical processes typical of semiconductors:\cite{Boyd,Meyer,JiWei}
\textit{\emph{bleaching}} and virtual two-photon processes, as represented
in Fig.~\ref{fig_two_photon}. Each of these processes excite different
electronic states. 

For sake of completeness, we give the formula for the optical susceptibility
due to three photon-absorption processes (i.e., the third-harmonic
generation, $e^{i3\omega t}$, neglected above): 
\begin{equation}
\chi_{3}^{(3\omega)}=\sigma_{0}{\cal W}_{i}\alpha\frac{3\pi v_{F}^{2}}{4\hbar\omega^{4}}\,.\label{eq:chi_third_harmonic}
\end{equation}

The transmittance of free-standing graphene for normal incidence is
obtained from 
\begin{equation}
{\cal T}=\frac{1}{\vert1+\chi(\omega)/(2\epsilon_{0}c)\vert^{2}}\,.\label{eq:transmittance}
\end{equation}
 Taking into account the non-linear corrections, we have 
\begin{equation}
\Re\chi(\omega)=\sigma_{0}(1-\beta_{\gamma})-\sigma_{0}\beta_{\omega}/2\,.\label{eq_chi_3_exact}
\end{equation}
 Since $\sigma_{0}/(2\epsilon_{0}c)=\pi\alpha$, the transmittance
of neutral graphene at zero temperature is 
\begin{equation}
{\cal T}\simeq1-\pi\alpha+\pi\alpha\beta_{\gamma}+\pi\alpha\beta_{\omega}/2\,,\label{eq_transmittance}
\end{equation}
where we have expanded Eq.~(\ref{eq:transmittance}) in the small
parameter $\alpha$. Higher-order terms are negligible except for
very high field intensities, that is, $\beta_{\gamma},\beta_{\omega}\gtrsim1$.
As expected, the non-linear contributions, $\beta_{\gamma}$ and $\beta_{\omega}$,
induce a higher transparency of graphene, which increases as ${\cal W}_{i}$
also
increases. {[}The imaginary part of $\chi(\omega)$, which we have
neglected, gives a small correction to Eq.~(\ref{eq_transmittance})---see
the top right panel of Fig.~\ref{fig_exact_conductivity} for the
magnitude of the imaginary part of $\Im\chi(\omega)$ in the rotating
wave approximation.{]}

It is important to realize that although the value of ${\cal W}_{i,\gamma}$,
coming from $\beta_{\gamma}$, suggests that the absorption of graphene
would saturate for moderate intensities, the much larger value of
${\cal W}_{i,\omega}$, coming from $\beta_{\omega}$, shows that
graphene still absorbs light due to virtual resonant two-photon processes,
even in the event of negligible \textit{\emph{bleaching}}. This is
possible because, as already noted, the two processes --\textit{\emph{bleaching}}
and the two-photon process-- excite different electronic states. If
light of broad spectral range is considered, instead of monochromatic
light, the analysis will be more complex than the one presented here.

\subsection{Approximate calculation of the non-linear susceptibility to all orders
in the intensity of the field within the RWA\label{sec_bleach_all}}

In the previous section we made a perturbative calculation of the
non-linear optical susceptibility of graphene, up to first order in
$\beta_{\gamma}$ and $\beta_{\omega}$. The exact calculation to
all orders  is not possible. However, an approximate
calculation of the non-linear susceptibility valid to all orders in
${\cal W}_{i}$ can be obtained using the rotating wave approximation
(RWA). Within the RWA, the solution of $X(t)=X$ is written as 
\begin{equation}
X(t)=xe^{-i\omega t}+ye^{i\omega t}\,.\label{eq_X_RWA}
\end{equation}
 Inserting $X(t)$ in Bloch's equations, we obtain 
\begin{eqnarray}
\dot{x}+(\gamma_{2}+i\epsilon-i\omega)x & = & i\frac{d_{\bm{k}}E_{0}}{2\omega}N\,,\\
\dot{y}+(\gamma_{2}+i\epsilon+i\omega)y & = & -i\frac{d_{\bm{k}}E_{0}}{2\omega}N\,,\\
\dot{N}+\gamma_{1}N & \approx & 4\frac{d_{\bm{k}}E_{0}}{2\omega}\Im(x-y)\,.
\end{eqnarray}
 The explicit solution of the above set of equations is obtained by
series resummation (assuming, for simplicity, $\gamma_{1}\approx\gamma_{2}\equiv\gamma$)
and reads: 
\begin{eqnarray}
x & = & iN\frac{d_{\bm{k}}E_{0}}{2\omega}\frac{\gamma+i(\omega-2v_{F}k)}{(\omega-2v_{F}k)^{2}+\gamma^{2}}\,,\label{eq_lx_RWA}\\
y & = & iN\frac{d_{\bm{k}}E_{0}}{2\omega}\frac{i(\omega+2v_{F}k)-\gamma}{(\omega+2v_{F}k)^{2}+\gamma^{2}}\,,\label{eq_ly_RWA}\\
N & = & \frac{n_{0}}{1+\delta(k)}\,,\label{eq_ln_RWA}
\end{eqnarray}
 where $\delta(k)$ is given by 
\begin{eqnarray}
\delta(k,\theta) & = & \tilde{\beta}\gamma^{2}\sin^{2}\theta\left(\frac{1}{(\omega-2v_{F}k)^{2}+\gamma^{2}}\right.\nonumber \\
 & + & \left.\frac{1}{(\omega+2v_{F}k)^{2}+\gamma^{2}}\right)\,,
\end{eqnarray}
 and $\tilde{\beta}$ is defined as $\tilde{\beta}=(8/3)\beta_{\gamma}$.

\begin{figure}[ht]
\begin{centering}
\includegraphics[clip,width=8cm]{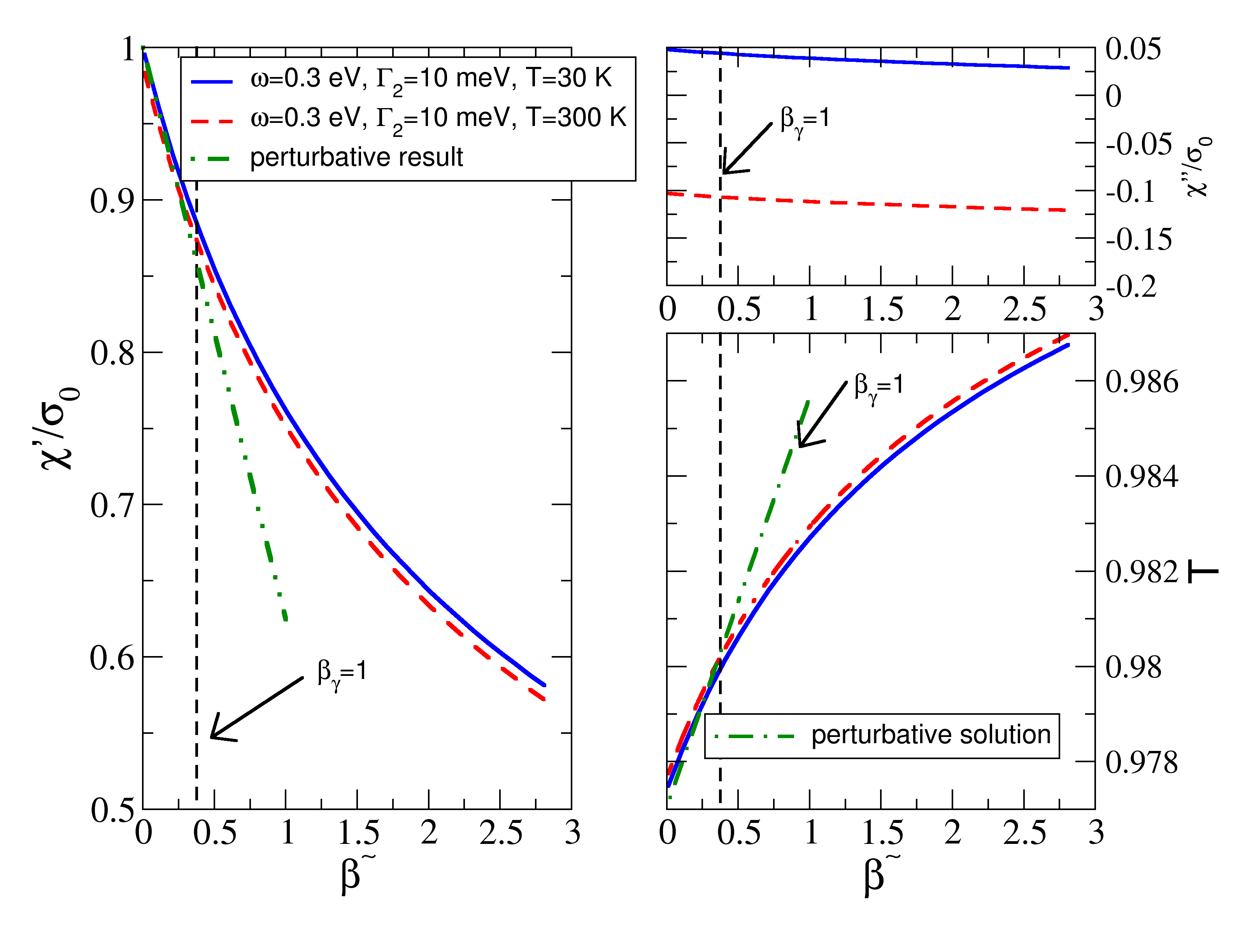} 
\par\end{centering}

\caption{\label{fig_exact_conductivity}Optical susceptibility of graphene
as function of $\tilde{\beta}$ within the RWA to all orders in ${\cal W}_{i}$.
In the left panel we plot the real part of the optical susceptibility
of graphene as function of the dimensionless parameter $\tilde{\beta}$
and in the right top panel we plot the imaginary part of the same
quantity ($\mu=10^{-3}$ eV). The transmittance, ${\cal T}$, of light
through graphene at normal incidence is plot in the right lower panel.
The vertical dashed line corresponds to $\beta_{\gamma}=1$. The results
are for $\hbar\omega=0.3$ eV and $\hbar\gamma=10$ meV and two different
temperatures, $T=30,300$ K. The dashed-dotted line is the perturbative
result given by Eqs.~(\ref{eq_chi_3_exact}) and (\ref{eq_transmittance}).}
\end{figure}

The calculation of the polarization follows the same procedure as
before, and using 
\begin{equation}
\int_{0}^{2\pi}d\theta\frac{\sin^{2}\theta}{1+a^{2}\sin^{2}\theta}=\frac{2\pi}{a^{2}}\frac{\sqrt{1+a^{2}}-1}{\sqrt{1+a^{2}}}\,,
\end{equation}
 we obtain 
\begin{equation}
P(t)=\frac{E_{0}}{2}e^{-i\omega t}[\chi'(\omega)+i\chi''(\omega)]+\mbox{c.c.}\,.\label{eq_Pol_b}
\end{equation}
 Considering zero temperature, with $\mu>0$ denoting the chemical
potential, the susceptibility reads 
\begin{equation}
\chi'(\omega)=\sigma_{0}\frac{2}{\pi\tilde{\beta}}\int_{2\mu/\hbar\gamma}^{\infty}dy\frac{\sqrt{1+\tilde{\beta}g(y)}-1}{\sqrt{1+\tilde{\beta}g(y)}}\,,\label{eq_real_bleach}
\end{equation}
 and 
\begin{eqnarray}
\chi''(\omega)=\frac{\omega}{\gamma}\chi'(\omega) & - & \sigma_{0}\frac{4}{\pi\tilde{\beta}}\frac{\omega}{\gamma}\int_{2\mu/\hbar\gamma}^{\infty}dy\frac{\sqrt{1+\tilde{\beta}g(y)}-1}{\sqrt{1+\tilde{\beta}g(y)}}\times\nonumber \\
 &  & \times\frac{y^{2}}{1+y^{2}+\omega^{2}/\gamma^{2}}\,,\label{eq_imag_bleach}
\end{eqnarray}
 where 
\begin{eqnarray}
g(y)=\frac{1}{(y-\omega/\gamma)^{2}+1}+\frac{1}{(y+\omega/\gamma)^{2}+1}\,.\nonumber \\
\end{eqnarray}

To zero order in $\tilde{\beta}$ we have the usual results (intra-band
contributions excluded): 
\begin{equation}
\chi'(\omega)=\sigma_{0}\left(1+\sum_{s=\pm1}\frac{s}{\pi}\arctan\frac{\hbar\omega-2\mu s}{\gamma\hbar}\right)\,,
\end{equation}
 and 
\begin{equation}
\chi''(\omega)=-\sigma_{0}\frac{1}{2\pi}\ln\frac{(2\mu+\hbar\omega)^{2}+\hbar\gamma^{2}}{(2\mu-\hbar\omega)^{2}+\hbar\gamma^{2}}\,.
\end{equation}
 To first order in $\tilde{\beta}$ and for neutral graphene at zero
temperature, we obtain for the real part of the susceptibility the
approximate result 
\begin{equation}
\chi'(\omega)\approx-\sigma_{0}\beta_{\gamma}-\sigma_{0}\beta_{\omega}\,.\label{eq_chi_3_RWA}
\end{equation}
 It is apparent from Eqs.~(\ref{eq_chi_3_exact}) and (\ref{eq_chi_3_RWA})
that the contribution coming from the \textit{\emph{bleaching}} process
is exact {[}first term in Eq.~(\ref{eq_chi_3_RWA}){]}, whereas the
contribution from virtual two-photon process is overestimated by
a factor of two in the RWA.

In Fig.~\ref{fig_exact_conductivity} we compare the perturbative
(dashed-dotted line) results given by Eqs.~(\ref{eq_chi_3_exact})
and (\ref{eq_transmittance}), left and right-bottom panels, respectively,
with  the RWA, valid for an arbitrary
value of ${\cal W}_{i}$ (we plot results for two different temperatures: $T=30$ K,
solid line, $T=300$ K dashed line). Clearly, the perturbative result
and the RWA calculation agree well up to $\beta_{\gamma}=1$, which
sets  perturbation theory validity limit. Beyond that
value a non-perturbative approach is necessary and one has to rely
on the RWA for drawing quantitative conclusions.

\section{An efficient graphene-based photodetector\label{sec_cavity}}

In this section we describe the absorption of light by a device composed
of optical cavities and a single graphene sheet. Only the linear optical
susceptibility will be considered, except when the light intensity
inside the cavity is of the order of ${\cal W}_{i,\gamma}$ (we remark
that for telecommunication devices and photodetectors this will hardly
be the case).

\subsection{Properties of a mirror and of an empty optical cavity\label{sec_cavity_empty}}

\begin{figure}[ht]
\begin{centering}
\includegraphics[clip,width=9cm]{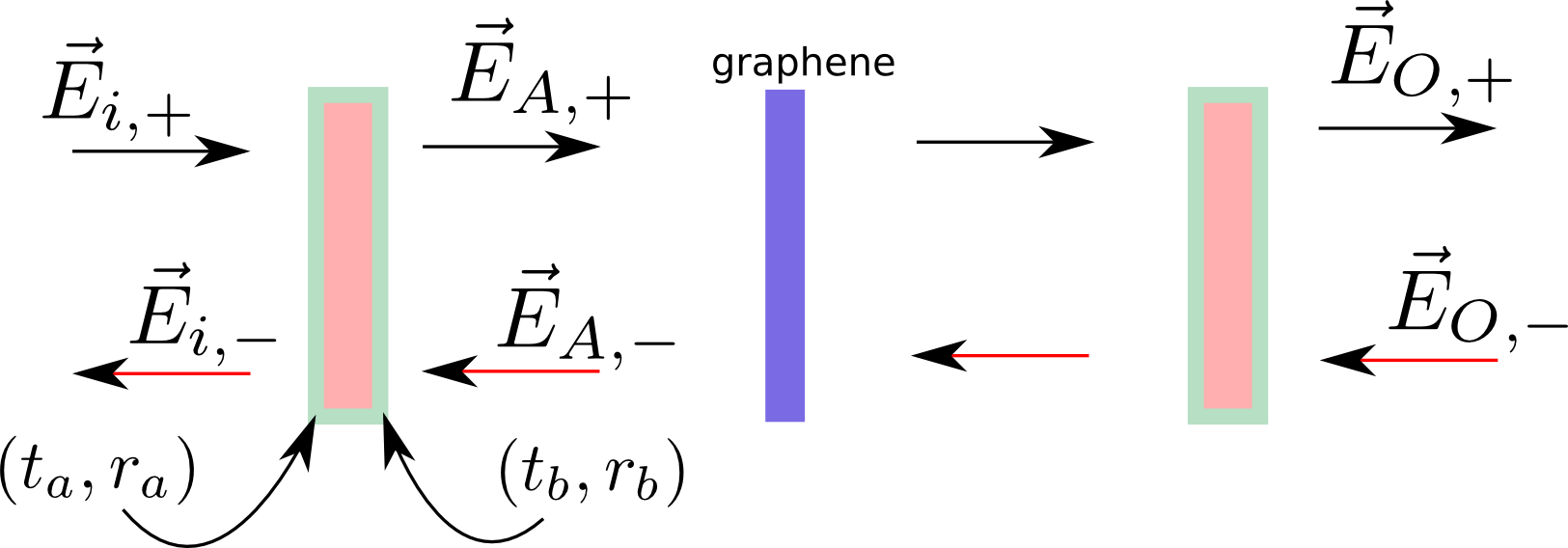} 
\par\end{centering}

\caption{\label{fig_cavity}Graphene inside an optical cavity. The cavity is
defined by two equal mirrors with reflectance amplitudes $r_{a}$ and $r_{b}$,
and transmittance amplitudes $t_{a}$ and $t_{b}$.}
\end{figure}

We start with the well-known case of an empty optical cavity. This
allow us to introduce important concepts and fix the notation.
The scattering matrix of a partially-silvered mirror is characterized
by two pairs of reflectance and transmittance coefficients; such a
mirror is shown in Fig.~\ref{fig_cavity}. 

The $S-$matrix relates the amplitudes of the incoming waves, $E_{i,+}$
and $E_{A,-}$, to the amplitudes of the outgoing waves, $E_{i,-}$
and $E_{A,+}$, according to

\begin{equation}
\left[\begin{array}{c}
E_{A,+}\\
E_{i,-}
\end{array}\right]=\left[\begin{array}{cc}
t_{a} & r_{b}\\
r_{a} & t_{b}
\end{array}\right]\left[\begin{array}{c}
E_{i,+}\\
E_{A,-}
\end{array}\right]\,.\label{eq_S_matrix}
\end{equation}
 On the other hand, the transfer matrix $M_{m}$ relates the fields
on the two sides of the mirror according to 
\begin{equation}
\left[\begin{array}{c}
E_{A,+}\\
E_{A,-}
\end{array}\right]=\frac{1}{t_{b}}\left[\begin{array}{cc}
t_{a}t_{b}-r_{a}r_{b} & r_{b}\\
-r_{a} & 1
\end{array}\right]\left[\begin{array}{c}
E_{i,+}\\
E_{i,-}
\end{array}\right]\,.\label{eq_M_matrix}
\end{equation}

Let us now discuss few properties obeyed by the reflectance and transmittance
amplitudes.\cite{Saleh} From the conservation of the energy flux
\begin{equation}
\vert E_{i,+}\vert^{2}-\vert E_{i,-}\vert^{2}=\vert E_{A,+}\vert^{2}-\vert E_{A,-}\vert^{2}\label{eq_energy_conserv}
\end{equation}
 we find 
\begin{eqnarray}
\vert t_{a}\vert^{2}+\vert r_{a}\vert^{2}=1\,,\\
\vert t_{b}\vert^{2}+\vert r_{b}\vert^{2}=1\,,\\
t_{a}r_{b}^{\ast}=-t_{b}^{\ast}r_{a}\,.
\end{eqnarray}
 From $t_{a}r_{b}^{\ast}=-t_{b}^{\ast}r_{a}$ we find 
\begin{equation}
\vert t_{a}\vert^{2}=\vert t_{b}\vert^{2}\frac{\vert r_{a}\vert^{2}}{\vert r_{b}\vert^{2}}\,,
\end{equation}
 which can be used to show that $\vert t_{a}\vert=\vert t_{b}\vert$
and $\vert r_{a}\vert=\vert r_{b}\vert$. Furthermore, the determinant
of the transfer matrix reads $\mbox{det\,}M_{m}=t_{a}/t_{b}$, implying
that $\vert\mbox{det\,}M_{m}\vert=\vert t_{a}/t_{b}\vert=1$. For
systems with inversion symmetry, we can write $t_{a}=t_{b}=t$ and
$r_{a}=r_{b}=r$. Thus, the $S-$matrix reads 
\begin{equation}
S=\left[\begin{array}{cc}
t & r\\
r & t
\end{array}\right]\,,\label{eq_S_matrix_symmetry}
\end{equation}
 and the transfer matrix can written as 
\begin{eqnarray}
M_{m} & = & \frac{1}{t}\left[\begin{array}{cc}
t^{2}-r^{2} & r\\
-r & 1
\end{array}\right]=\left[\begin{array}{cc}
t+\vert r\vert^{2}/t^{\ast} & r/t\\
-r/t & 1/t
\end{array}\right]\nonumber \\
 & = & \left[\begin{array}{cc}
1/t^{\ast} & r/t\\
r^{\ast}/t^{\ast} & 1/t
\end{array}\right]\,.\label{eq_M_matrix_symmetry}
\end{eqnarray}
 Writing $r=\vert r\vert e^{i\alpha_{r}}$ and $t=\vert t\vert e^{i\alpha_{t}}$,
the relation $tr^{\ast}=-t^{\ast}r$ implies that $e^{i2(\alpha_{r}-\alpha_{t})}=-1$,
that is, $\alpha_{r}=\alpha_{t}\pm\pi/2$. Using these last relations,
$M_{m}$ can be written as 
\begin{equation}
M_{m}=\frac{1}{t}\left[\begin{array}{cc}
-1 & -\vert r\vert\\
\vert r\vert & 1
\end{array}\right]\,.
\end{equation}

If  two of these mirrors are separated by a distance $L$ we have
to define the transfer matrix associated with the free propagation
from the first to the second mirror. Since 
\begin{eqnarray}
E_{+}(x,t) & = & E_{+}e^{i(kx-\omega t)}\,,\\
E_{-}(x,t) & = & E_{-}e^{-i(kx+\omega t)}\,,
\end{eqnarray}
 then at a distance $L$ to the right the $E_{+}(x+L,t)$ has an extra
phase of $e^{ikL}$ whereas the $E_{-}(x+L,t)$ has an extra phase
of $e^{-ikL}$. Thus we have 
\begin{equation}
\left[\begin{array}{c}
E_{A,+,L}\\
E_{A,-,L}
\end{array}\right]=\left[\begin{array}{cc}
e^{ikL} & 0\\
0 & e^{-ikL}
\end{array}\right]\left[\begin{array}{c}
E_{A,+}\\
E_{A,-}
\end{array}\right]\,,
\end{equation}
 or 
\begin{equation}
\left[\begin{array}{c}
E_{A,+}\\
E_{A,-}
\end{array}\right]=\left[\begin{array}{cc}
e^{-ikL} & 0\\
0 & e^{ikL}
\end{array}\right]\left[\begin{array}{c}
E_{A,+,L}\\
E_{A,-,L}
\end{array}\right]\,,\label{eq_free_m}
\end{equation}
 where $E_{A,+/-,L}$ represents the amplitude of the forward/backward
propagating field at the right end of the cavity and the matrix 
\begin{equation}
M_{f}(L)=\left[\begin{array}{cc}
e^{-ikL} & 0\\
0 & e^{ikL}
\end{array}\right]
\end{equation}
 defines the free propagation to the right. Then, the transmitted
field through two mirrors at a distance $L$ from each other follows from 
\begin{equation}
\left[\begin{array}{c}
E_{i,+}\\
E_{i,-}
\end{array}\right]=M_{m}\cdot M_{f}(L)\cdot M_{m}\left[\begin{array}{c}
E_{O,+}\\
E_{O,-}
\end{array}\right]\,.
\end{equation}
 Explicitly, we have 
\begin{equation}
E_{i,+}=\frac{e^{-ikL}}{(t^{\ast})^{2}}[1-(r^{\ast})^{2}e^{2ikL}]E_{O,+}\,,
\end{equation}
 writing $r^{\ast}=\vert r\vert^{-i\alpha_{r}}$ we obtain 
\begin{equation}
{\cal T}\equiv\frac{\vert E_{O,+}\vert^{2}}{\vert E_{i,+}\vert^{2}}=\frac{\vert t\vert^{4}}{\vert t\vert^{4}+4\vert r\vert^{2}\sin^{2}(kL)}\,,\label{eq_transmission}
\end{equation}
since $\alpha_{r}=\pi$, as implied by Fresnel equations. Thus we have
perfect transmission for $kL=n\pi$ or $\lambda=2L/n$, with $n=1,2,3,\ldots$.
The longest wavelength for which perfect transmission is possible
is $\lambda=2L$. It is straightforward to show that the transmission
is strongly suppressed for other choices of $kL$. For instance, when
$kL$ is a multiple of $\pi/2$, we have 
\begin{equation}
{\cal T}=\frac{\vert t\vert^{4}}{\vert t\vert^{4}+4\vert r\vert^{2}}\approx\frac{\vert t\vert^{4}}{4\vert r\vert^{2}}\ll1\,,
\end{equation}
In the above, we have admitted high-quality mirrors, $\vert r\vert\gg\vert t\vert$,
to simplify the denominator. 

The introduction of a graphene sheet inside the cavity leads to light
absorption and the relation (\ref{eq_transmission}) is modified.
In the following section, we demonstrate how to explore the physics
of an optical cavity to devise an efficient graphene-based  photodetector.

\subsection{Graphene in an optical cavity\label{sec_cavity_simple}}

We describe the transmission of light through a graphene sheet inside
an optical cavity taking into account the linear optical-susceptibility
of graphene. 
\begin{figure}[ht]
\begin{centering}
\includegraphics[clip,width=9cm]{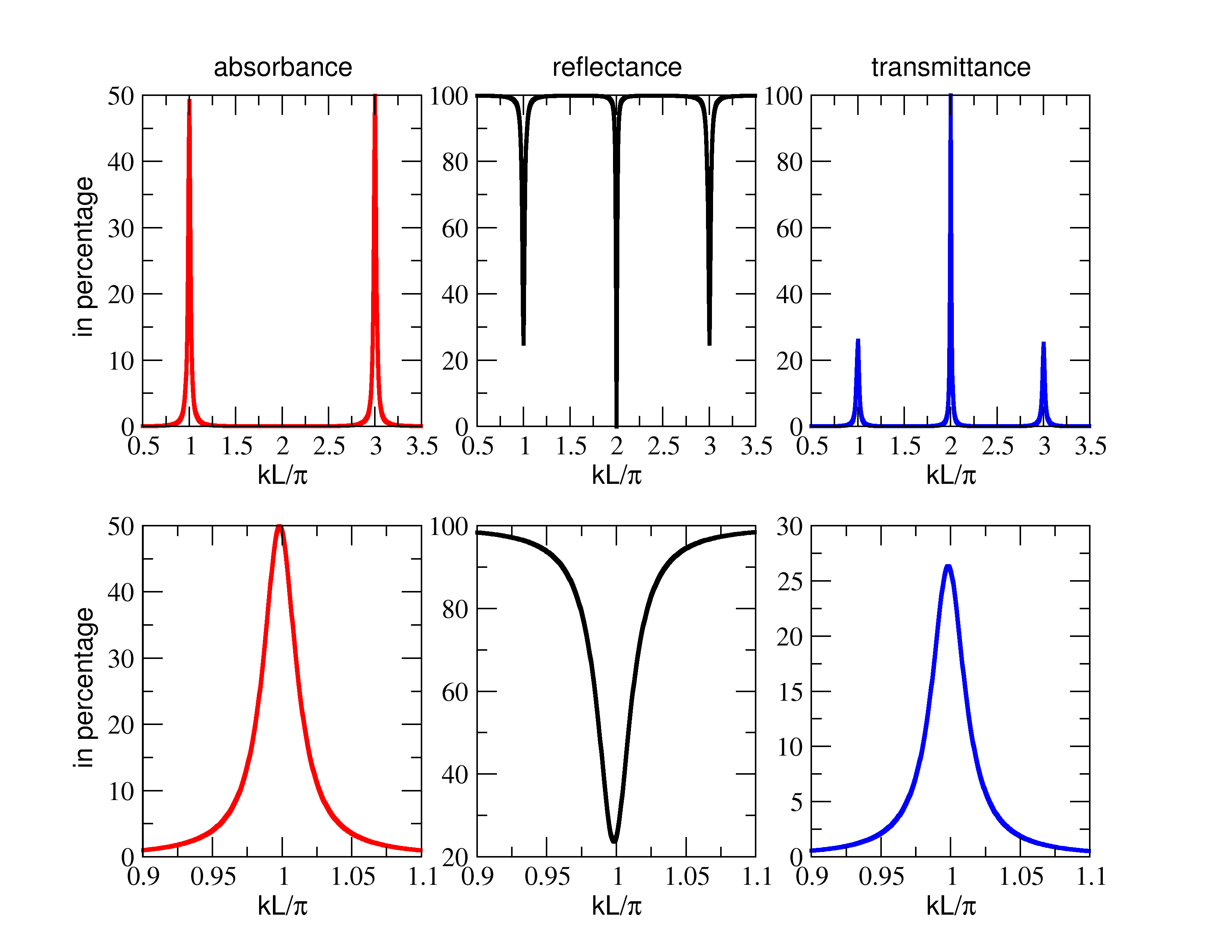} 
\par\end{centering}

\caption{\label{fig_finesse}Transmission spectrum for a Fabry-P\'erot cavity
with graphene $L/2$. The mirror transmittance is
$t^{2}=0.045$. If we tale $\lambda=1000$ nm, then the range $kL/\pi$
in the lower panels spans the wavelengths from $\lambda=900$ nm to
$\lambda=1100$ nm. We have taken $\hbar\gamma=7$ meV and $T=300$
K. Regarding the choice of $\lambda$, we note that the first HeNe
laser was working at the spectral wavelength of 1150~nm, that is
in the infrared.}
\end{figure}

We write the transfer matrix of graphene as\cite{EOM} 
\begin{equation}
M_{g}=\left[\begin{array}{cc}
1+\eta & \eta\\
-\eta & 1-\eta
\end{array}\right]\,,
\end{equation}
where $2\eta=Z_{0}\chi(\omega)$, with $\eta=\eta'+i\eta''$, and
$Z_{0}\simeq376.7$~$\Omega$ is the vacuum impedance. For neutral
graphene at zero temperature $\eta$ is essentially a real number
for frequencies below the visible spectral range. The transmission
through the cavity with graphene at position $x_{g}$ and the second
mirror at position $x=L$ follows from 
\begin{equation}
\left[\begin{array}{c}
E_{i,+}\\
E_{i,-}
\end{array}\right]=M_{m}\cdot M_{f}(x_{g})\cdot M_{g}\cdot M_{f}(L-x_{g})\cdot M_{m}\left[\begin{array}{c}
E_{O,+}\\
E_{O,-}
\end{array}\right]\,.\label{eq_M_for_single_cavity}
\end{equation}
The matrix 
\begin{equation}
M=M_{m}\cdot M_{f}(x_{g})\cdot M_{g}\cdot M_{f}(L-x_{g})\cdot M_{m}
\end{equation}
is the full transfer matrix of the device. The transmittance and the
reflectance are defined as 
\begin{eqnarray}
{\cal T} & = & \frac{1}{\vert M_{11}\vert^{2}}\,,\\
{\cal R} & = & \left\vert \frac{M_{21}}{M_{11}}\right\vert ^{2}\,,
\end{eqnarray}
respectively, and $M_{11}$ and $M_{12}$ denote the matrix elements
of $M$. We note that ${\cal R}\ne1-{\cal T}$ due to absorption by
graphene (we are assuming lossless mirrors). The absorbance is defined
as ${\cal A}=1-{\cal R}-{\cal T}.$

For $x_{g}=L/2$ we have 
\begin{eqnarray}
{\cal T} & = & \frac{\vert t\vert^{4}}{\vert1+\eta-2\eta\vert r\vert e^{ikL}-(1-\eta)\vert r\vert^{2}e^{2ikL}\vert^{2}}\,,\label{eq_T_cav_sing}\\
{\cal R} & = & \frac{\vert(1+\eta)|r|-\eta(1+|r|^{2})e^{ikL}-(1-\eta)|r|e^{2ikL}\vert^{2}}{\vert1+\eta-2\eta\vert r\vert e^{ikL}-(1-\eta)\vert r\vert^{2}e^{2ikL}\vert^{2}}\,.
\end{eqnarray}
In the limit $r\rightarrow0$ we recover the well known result, ${\cal T}=1/\vert1+\eta\vert^{2}$
{[}see also Eq.~(\ref{eq:transmittance}){]} 
and ${\cal R}=\vert\eta\vert^{2}/\vert1+\eta\vert^{2}$.\cite{stauberFullSigma} 

\begin{figure}[ht]
\begin{centering}
\includegraphics[clip,width=9cm]{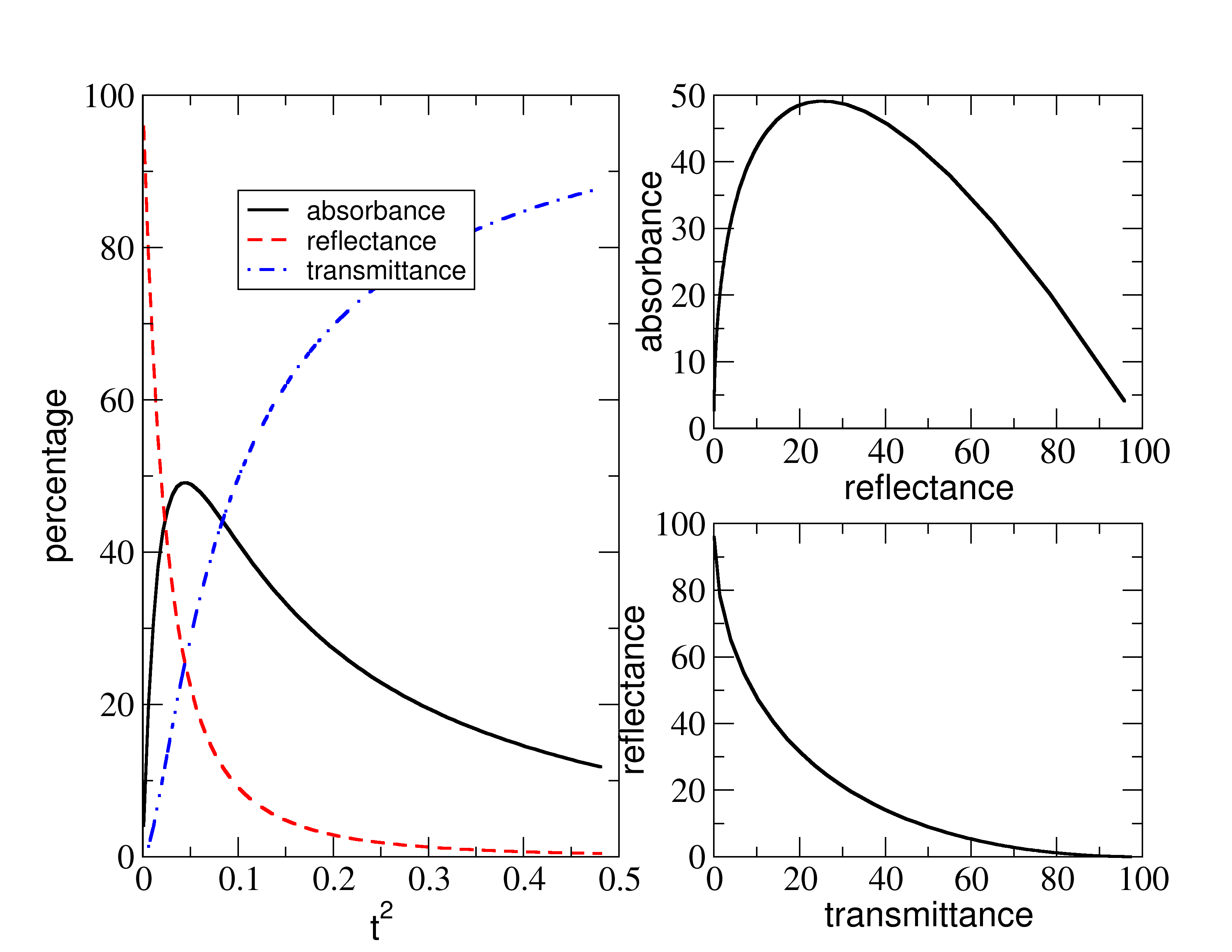} 
\par\end{centering}

\caption{\label{fig_Absorvance_func_t_one_cavity}Absorbance of graphene for
the single cavity system (see Fig.~\ref{fig_cavity}). Left: Absorbance
as function of mirror transmittance $t^{2}\in[0,0.5]$. Top right:
Absorbance of graphene as function of reflectance. Bottom right: reflectance
of graphene as function of transmittance. The calculations assumed
$T=300$ K.}
\end{figure}

The effect of graphene in the cavity is to reduce the intensity of
the odd-orders ($n=1,3,5,\ldots$) of perfect transmission in the
otherwise perfect cavity. From Fig.~\ref{fig_finesse} it is clear
that the reduction of transmission of the odd orders is divided between
reflection and absorption, the latter taking the majority of the incoming
power. The transmission is still unity for  even-orders ($n=2,4,6,\ldots$).
In Fig.~\ref{fig_Absorvance_func_t_one_cavity} we show the dependence
of the absorbed power as function of the transmittance $t^{2}$ of
a mirror. Clearly, this dependence is not monotonous, displaying a
maximum around $t^{2}\simeq0.045$. An analytical expression for the
value of $t^{2}$ for which the absorption is maximum can be readily
obtained from Eq.~(\ref{eq_T_cav_sing}). 

\subsection{Double optical cavity\label{sec_cavity_double}}

\begin{figure}[ht]
\begin{centering}
\includegraphics[clip,width=9cm]{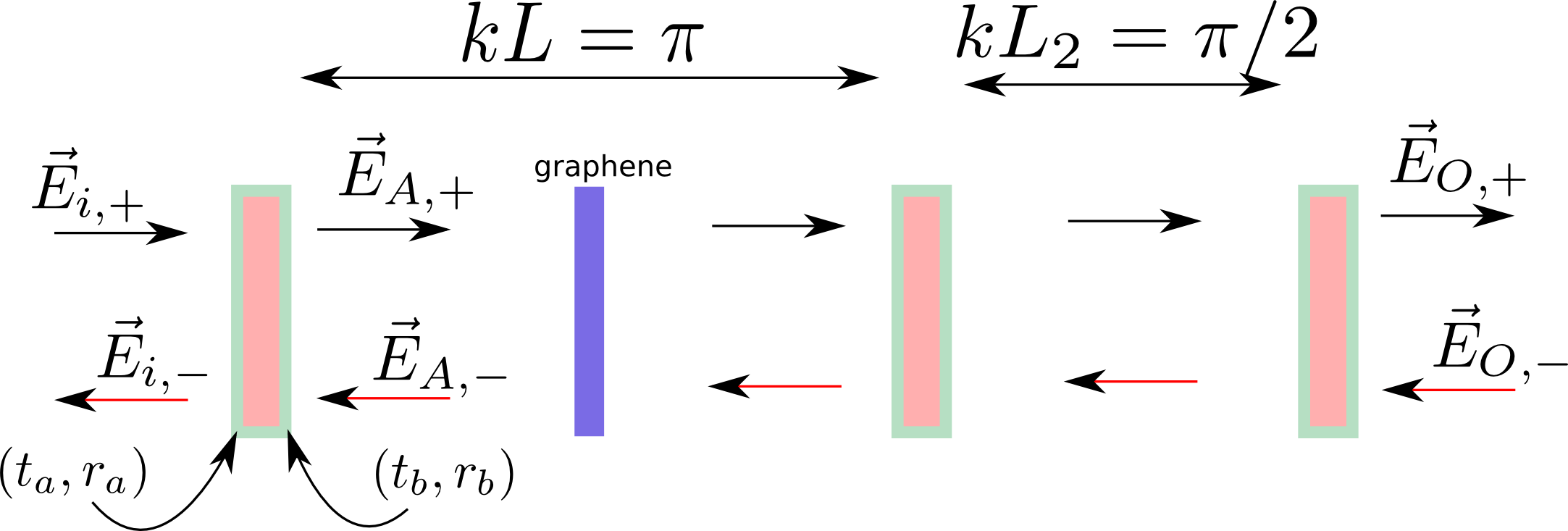} 
\par\end{centering}

\caption{\label{fig_two_cavities}The double cavity system: Light impinges
from the left-hand side. The second cavity has half the length of
the first cavity, $L_{2}=L/2$.}
\end{figure}

The goal of the present section is to discuss a device able to enhance
light absorption relatively to the single cavity system discussed
above. To that end we consider a graphene sheet inside an optical
cavity of length $L$ (as in the previous section) followed by an
empty quarter-wavelength cavity, as represented in Fig.~\ref{fig_two_cavities}.

Placing graphene at the middle of the first cavity and choosing its
length $L$ such that $L=\lambda/2$, where $\lambda$ is the wavelength
of the light, we expect that graphene will present an enhanced absorption
at this wavelength, at least for a cavity with a high finesse. This
intuitive picture is developed from considering, as a rough approximation,
the formation of standing waves within the cavity having their maximum
amplitudes at the center of the cavity. This picture in confirmed
by simulations, as shown in Fig.~\ref{fig_finesse}. 
Also from Fig.~\ref{fig_Absorvance_func_t_one_cavity}
it is clear that there is an optimal value of $t^{2}$ for which the
absorption can be as high as 50\% (the quantitative results are robust
for small changes graphene's position  relatively to the center
of the cavity).

\begin{figure}[ht]
\begin{centering}
\includegraphics[clip,width=9cm]{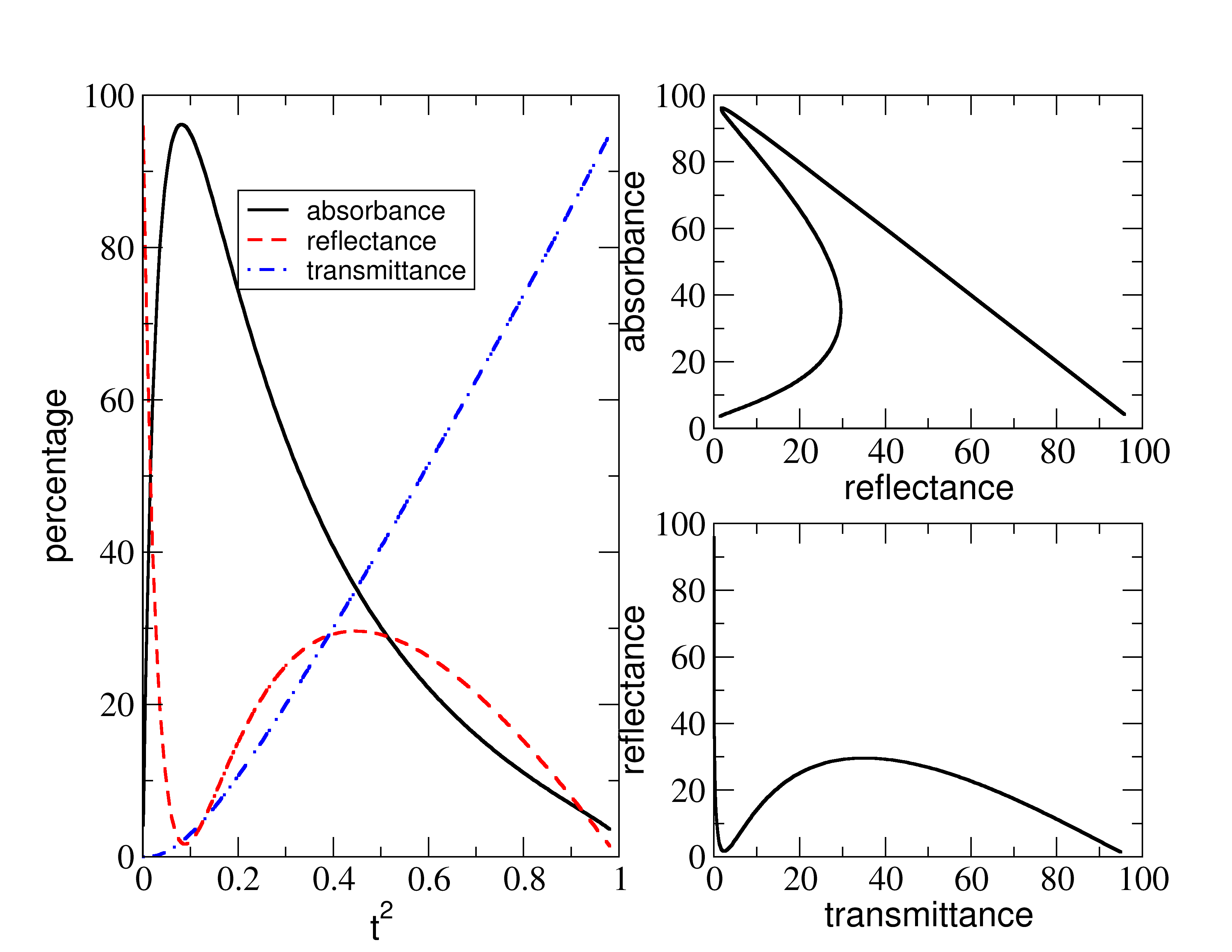} 
\par\end{centering}

\caption{\label{fig_Absorvance_func_t_two_cavities}Absorbance of graphene
for the double cavity setup (see Fig.~\ref{fig_two_cavities}).
Left: Absorbance as function of mirror transmittance $t^{2}\in[0,1]$.
Top right: Absorbance as function of reflectance. Bottom right: Reflectance
as function of transmittance. The calculations assumed $T=300$ K.}
\end{figure}

\begin{figure}[ht]
\begin{centering}
\includegraphics[clip,width=9cm]{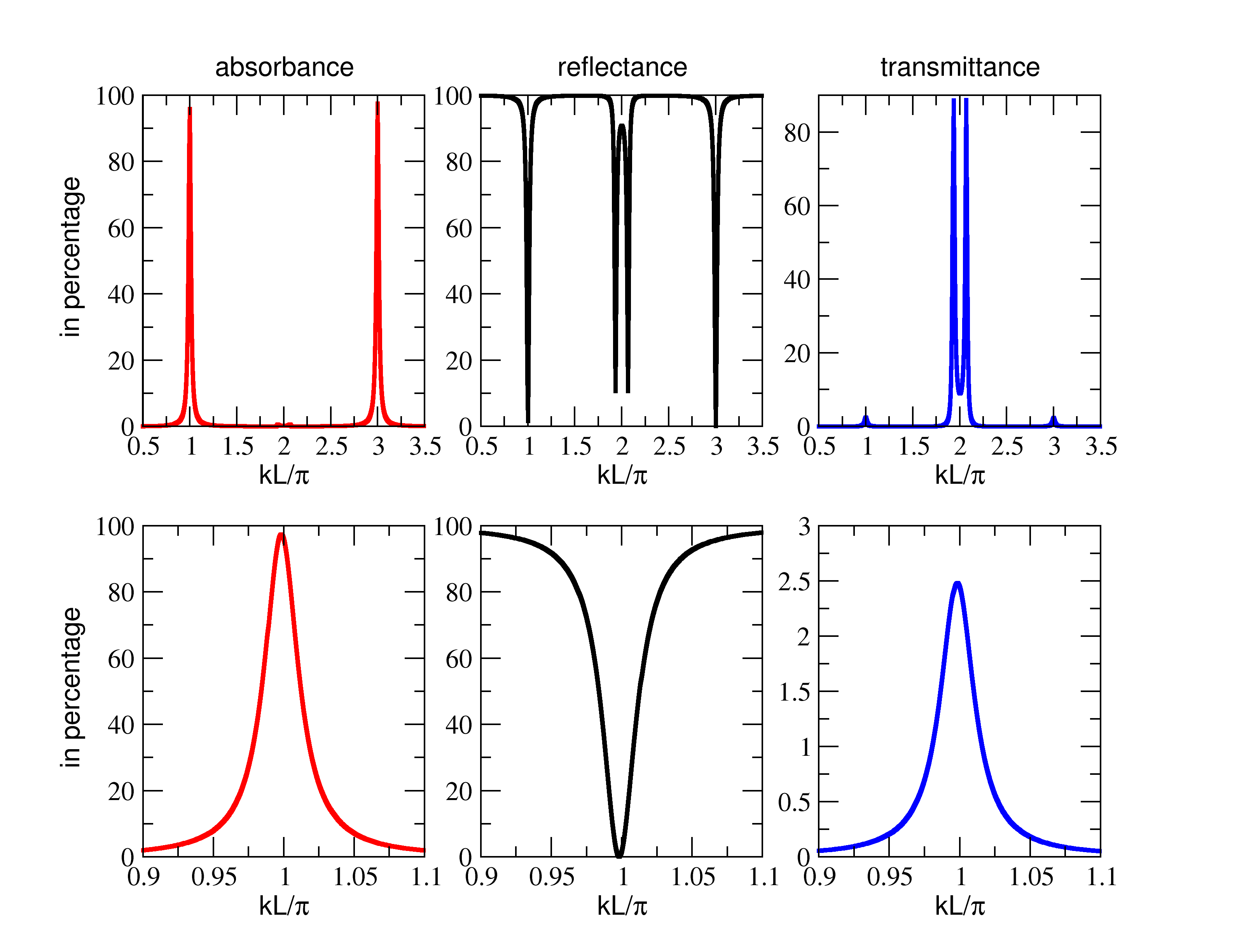} 
\par\end{centering}

\caption{\label{fig_Absorvance_func_w}Absorbance as function of $kL/\pi=2L/\lambda$
for the double cavity system ($t^{2}=0.09$). If we consider $\lambda=1000$
nm, then the range of $kL/\pi$ in the lower panels spans the wavelengths
from $\lambda=900$ nm to $\lambda=1100$ nm.}
\end{figure}

The absorbed intensity can be pushed up to $\simeq100\%$ by building
an optical cavity containing graphene, followed by an empty quarter-wavelength
cavity, that is, a cavity with length $L_{2}$ such that $L_{2}=\lambda/4$.
This setup leads to an enhancement of the absorption, which is about
twice as large as that of the single cavity setup, as can be seen
in the left panel of Fig.~\ref{fig_Absorvance_func_t_two_cavities}.
The dependence of the absorption on the wavelength is 
shown in Fig.~\ref{fig_Absorvance_func_w}.

A physical qualitative argument for the absorption enhancement effect
in a double cavity is reminiscent of a quantum particle in a box with
a permeable wall. Let us consider first a box with origin at $x=0$
and length $L+L/2$. A wall is located at $x=L$. If the wall is impermeable,
the fundamental mode of the the first box is $\lambda=2L$ and that
of the second box is $\lambda=L$. If the wall becomes permeable the
two modes hybridize and in the ground state the probability density
grows in the first box at the expenses of the probability density
in the second box. Translating this into our problem, the quarter
wavelength cavity interference between the wave reflected by the third
mirror and the forward propagating wave effectively suppresses the
transmission at wavelengths $\lambda=2L$, forcing the photon to spend
more time in the first cavity, and thus increasing the absorption
by the graphene sheet.

It is worth stressing that the maximum of absorption takes place for
cavities with $\vert t\vert^{2}\simeq0.1$, a convenient figure from
the point of view of micro-fabrication since not much effort has to
be put on building highly reflective mirrors.

From a theoretical point of view, the calculation of the properties
of the two coupled cavities follows from the transfer matrix method,
reviewed in \ref{sec_cavity_simple}. As in Eq.~(\ref{eq_M_for_single_cavity}),
the incoming and outgoing field amplitudes are related as 
\begin{equation}
\left[\begin{array}{c}
E_{i,+}\\
E_{i,-}
\end{array}\right]=M_{{\rm cav}}\left[\begin{array}{c}
E_{O,+}\\
E_{O,-}
\end{array}\right]\,,\label{eq_M_for_two_cavities}
\end{equation}
 where the transfer matrix of the two cavities is given by 
\begin{equation}
M_{{\rm cav}}=M_{m}\cdot M_{f}(x_{g})\cdot M_{g}\cdot M_{f}(L-x_{g})\cdot M_{m}\cdot M_{f}(L_{2})\cdot M_{m}\,.
\end{equation}
 In Figs.~\ref{fig_Absorvance_func_t_two_cavities} and \ref{fig_Absorvance_func_w}
we have considered $x_{g}=L/2$. In this case the field amplitudes
for $\lambda=(2n+1)L/2$ (with $n=0,1,2,\ldots$) have a maximum at
the center of the cavity. As for the case of the single cavity, it
is possible to derive analytical expressions for both ${\cal T}$
and ${\cal R}$; we obtain 
\begin{equation}
\mathcal{T}=\frac{|t|^{6}}{\left|1+\eta+\Lambda_{1}e^{ikL}+\Lambda_{2}e^{2ikL}+\Lambda_{3}e^{3ikL}\right|^{2}}\,.
\end{equation}
In the above, $\Lambda_{1}=-[\eta(2+|r|)+|r|]|r|$, $\Lambda_{2}=[\eta(1+|r|+|r|^{2})-|r|]|r|$
and $\Lambda_{3}=(1-\eta)|r|^{2}$, and 
\begin{equation}
\mathcal{R}=\frac{\mathcal{T}}{|t|^{6}}\left|(1+\eta)|r|+\Delta_{1}e^{ikL}+\Delta_{2}e^{2ikL}+\Delta_{3}e^{3ikL}\right|^{2}\,,
\end{equation}
with $\Delta_{1}=-[\eta(1+|r|^{2}+|r|^{3})+|r|^{3}]$, $\Delta_{2}=(\eta-1+2|r|\eta)|r|$
and $\Delta_{3}=(1-\eta)|r|$. As before, taking the limit $r\rightarrow0$
we recover the well known values for ${\cal T}$ and ${\cal R}$ in
the absence of mirrors.

It is interesting to note the hysteresis in the absorbance as function
of the reflected power ---some values of the reflectance admit three
possible absorbances; see top right panel of Fig.~\ref{fig_Absorvance_func_t_two_cavities}.
Each point on the ${\cal A}$ versus $\mathcal{R}$ curve corresponds
to a given value of the mirror transmittance, $\vert t\vert^{2}$,
which therefore can be viewed as an external driving field. 

Finally, we note that about 100\% absorption can also be obtained
for a single cavity if the second mirror reflects 100\% of the 
impinging light. The curves ${\cal A}$, ${\cal R}$, and ${\cal T}$ as
function of $t^{2}$ are different in this case, however, from those given above; 
in particular,
${\cal A}$ shows no hysteresis (see note after {\it Conclusions} section).\cite{micro}

\section{Conclusions\label{sec_conclu}}

In the present article, we have developed the theory a graphene-based photodetector
with nearly 100\% efficiency for photon frequencies around a predefined
value. The proposed setup is general and should work in a vast spectral
range. 

We have also clarified the role of the non-linear optical susceptibility
in determining the properties of the cavity-graphene system. We have
shown that  non-linear terms are irrelevant for moderate light
intensities. 

The most efficient photodetector is built from combining
a half-wavelength cavity (size $L=\lambda/2$) followed by a second quarter-wave
length cavity (size $L_{2}=\lambda/4$). This system improves the
absorption by a factor of two relatively to the single cavity. As
noted above, the two-cavities photodetector has about the same absorbance
as a single cavity with the second mirror having zero transmittance
and the first one having $t^{2}\simeq0.9$. 

If real-time control of the mirrors transmittance is feasible, then it 
will be 
possible to obtain hysteresis in the absorbance for the same reflectance
value. Whether this can be used as an optical logical gate is so far
unclear and will be left for future research.

Although in the schematic figures for the cavities graphene appears
floating in the air, in practical terms it will be deposited on a
dielectric. The mirrors can also be made of dielectric materials,
the so called Bragg mirrors. There are computer
codes for simulating mirrors with the prescribed optimal value of
$t^{2}$ given in the text. The setup itself
 can be built by micro-fabrication
using standard techniques. 

\textit{Note: during the final state of writing, we became
aware of an experimental paper, entitled ``Microcavity-integrated
graphene photodetector''. In that work a single cavity detector
has been built.\cite{micro} Our theory is a full analytical account
of the physics of two similar devices, one of them having the same
theoretical efficiency as the device of that paper but a different
qualitative response as function of the amplitude $t^{2}$.}

A.F. acknowledges FCT Grant No. SFRH/BPD/65600/2009. N.M.R.P. and
R.M.R. acknowledge Fundos FEDER, through the Programa Operacional
Factores de Competitividade - COMPETE and by FCT under Project No.
Past-C/FIS/UI0607/2011. T.S. acknowledges FCT Grant No. PTDC/FIS/101434/2008
and FIS2010-21883-C02-02 (MICINN).

\end{document}